\documentclass{iaus}
\usepackage{graphicx}

\title[Massive stars and their surroundings] 
{Interaction of massive stars with their surroundings}

\author[Gerhard Hensler]   
{Gerhard Hensler
}

\affiliation{Institute of Astronomy, University of Vienna,
Tuerkenschanzstr. 17, 1180 Vienna, Austria \\
email: hensler@astro.univie.ac.at}

\pubyear{2008}
\volume{252}  
\pagerange{119--126}
\date{?? and in revised form ??}
\jname{The art of modelling stars in the 21$^{st}$ century}
\editors{L. Deng \& K.L. Chan, eds.}
\begin{document}

\def\HI{H{\sc i} }
\def\HII{H{\sc ii} }
\def\Ha{H$\alpha$ }
\def\OIII{O[{\sc iii}] }
\def\Msun{M$_{\odot}$ }
\def\eps{$\epsilon$}
\def\OH{{12\-+log(O/H)} }
\def\lNO{log(N/O) }
\def\cd{chemo-dynamical }
\def \rmaa {Rev.Mex.\allowbreak  Astron.\allowbreak  Astrofis.}
\def\AA{A{\rm \&}A}

\maketitle

\begin{abstract}
Due to their short lifetimes but their enormous energy release
in all stages of their lives massive stars are the major engines
for the comic matter circuit. They affect not only their close
environment but are also responsible to drive mass flows on galactic scales.
Recent 2D models of radiation-driven and wind-blown \HII regions are
summarized which explore the impact of massive stars to the interstellar medium
but find surprisingly small energy transfer efficiencies while an observable
Carbon self-enrichment in the Wolf-Rayet phase is detected in the warm ionized
gas.
Finally, the focus is set on state-of-the-art modelling of \HII regions
and its present weaknesses with respect to uncertainties and simplifications
but on a perspective of the requested art of their modelling in the 21$^{st}$
century.
\end{abstract}

\firstsection 
\section{Introduction}

Massive stars play a crucial role in the evolution of galaxies, as they are
the primary source of metals, and they dominate the turbulent energy
input into the interstellar medium (ISM) by their massive
and fast stellar winds, by the ultraviolet radiation, and by supernova explosions.
The radiation field of these stars, at first, photo-dissociates the ambient
molecular gas and forms a so-called photo-dissociation region (PDR) of neutral
hydrogen. Subsequently, the Lyman continuum photons of the star ionize
the \HI gas and produce a \HII region that expands into the neutral ambient medium.

As these stars have short lifetimes of only a few million years,
\HII regions indicate the sites of star formation (SF) and are targets to measure
the current SF rate in a galaxy. Furthermore, the
emission line spectrum produced by the ionized gas allows the accurate determination
of the current chemical composition of the gas in a galaxy.
Although the physical processes of the line excitation are quite well understood and
accurate atomic data are available, so that the spectral analysis of \HII regions
(see e.g. \cite{stas79,evdo87}) serves as an essential tool to study
the evolution of galaxies, their reliability as diagnostic tool have also to be
studied with particular emphasis e.g.\ to temperature fluctuations
(\cite{peim67,stas02}) and line excitations.

The simple concept of a uniform medium in ionization equilibrium with the
radiation from a massive star (the Str\"omgren sphere) is successful
in describing several global features of \HII regions and allows
to model the emission line spectrum to a first-order reliability.
Since it has long been realized that \HII regions are also excerting
complex dynamics to the ISM, dynamical modelling of
\HII regions caused purely by the energy deposit of the stellar
radiation field has therefore been started already long ago
(see e.g \cite{york86} and references therein)
providing a first insight into the formation of dynamical structures.
In addition, also an expanding stellar wind bubble (SWB) with a
constant wind power but neglecting its dynamics can be analytically
described for the adiabatic phase as the self-similar evolution
by Sedov.

\begin{figure}
\begin{center}
 \includegraphics[width=3.0in]{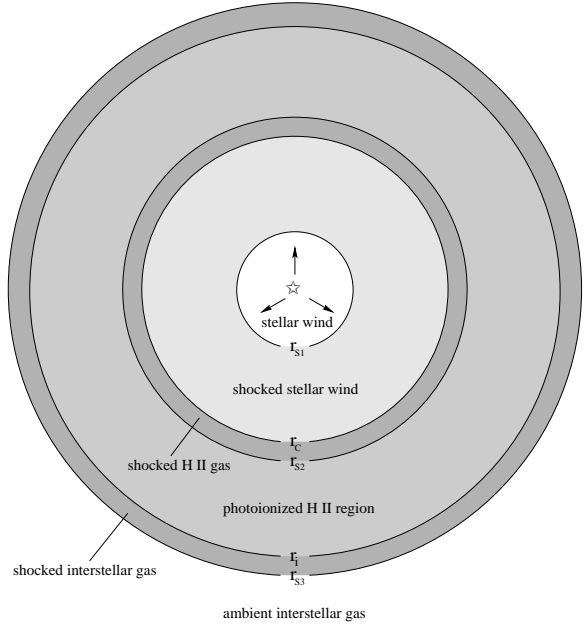}
  \caption{Schematic structure of a SWB as derived by \cite{weav77}:
rs1 marks the position of
the reverse shock, rc the contact discontinuity, rs2 the forward shock of the
stellar wind bubble, ri the ionization front, and rs3 the forward shock of the
\HII region expansion.}
\end{center}
\end{figure}

In order to approach reality of how a radiation-driven and wind-blown
SWB around a massive star interacts with its surroundings one has
to take the kinetic energy of the wind and its dissipation into account and
has to consider the ionization of the neutral environmental gas where it
still exists. Analytically, this can be allowed for in spherical symmetry
under the simplifying assumptions of a point source
of constant strong wind that interacts with a homogeneous ambient
ISM as derived by \cite{weav77}. This leads to
a clear stratification of the surrounding bubble (see fig. 1) from inside
out: freely expanding stellar wind, reverse shock, shocked stellar wind,
contact discontinuity, SWB forward shock, photo-ionized \HII region,
ionization front, forward-shocked ambient gas.

Although the analytical and semi-analytical solutions for the evolution of
SWBs have been improved over the years as well as the numerical simulations
have been done with increasing complexity, like e.g.\ 2D calculations of SWBs
(see e.g. by \cite{rozy85} and a series of papers)
and/or combined 1D radiation-hydrodynamical models of \HII regions coupled
with the dynamical SWB (for references see Freyer et al. 2003)
a variety of physical effects remains to be included in order to
achieve a better agreement of models and observations e.g. with regard to the
evolution of the hot phase in bubbles (\cite{mclo00,chu00}).

\section{SWB Models}

To improve the insight into the evolution of radiation-driven + wind-blown
bubbles around massive stars, we have performed a series of
radiation-hydrodynamical simulations with a 2D cylindrical-symmetric
nested-grid scheme for stars of masses 15 \Msun (Kroeger et al. in prep.),
35 \Msun (\cite{frey06}), 60 \Msun (\cite{frey03}),
and 85 \Msun (\cite{kroe07}).
The main issues of these models can be summarized as follows:

\begin{figure}
\begin{center}
 \includegraphics[width=5.0in]{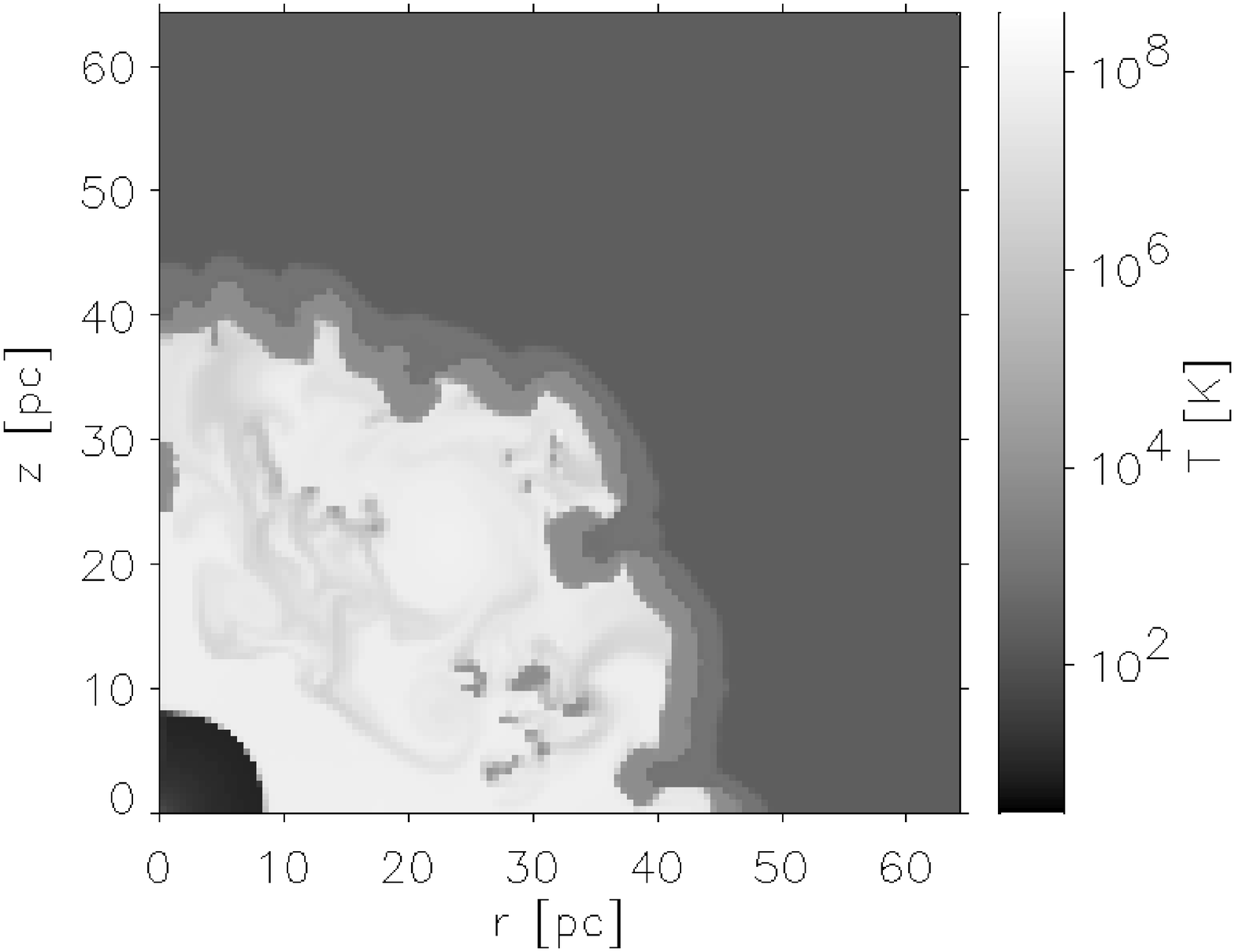}
  \caption{Temperature distriution of stellar wind bubble + \HII region
surrounding a 60 \Msun star at an age of 3.3 Myr
(\cite{frey03}). }
\end{center}
\end{figure}

\noindent
1) The \HII regions formed around the SWBs have complex structures
mainly affected by dynamical processes, such as shell instabilities,
vortices, mixing effects, etc. (see fig. 2)

\noindent
2) The stronger the wind the higher the compresssion of the SWB-surrounding
\HII region (\cite{hens08}), what means that the \HII rgion does not increase
according to the larger ionizing flux.

\noindent
3) Finger-like and spiky structures of different densities and temperatures
are formed in the photo-ionized region (\cite{frey03}).

\noindent
4) The regions contributing to the \HII region emission line spectrum are not solely
limited to the photo-ionized shell around the SWB but also form from
photo-evaporated gas at the trailing surface of the SWB shock front
(\cite{hens08}; and see fig. 2).

\noindent
5) Because dispersion of this cooler photo-evaporated gas into the hot SWB
leads to mixing also the stellar material expelled by the wind has to
emerge partly in the \HII region spectra.

\noindent
6) As a consequence the metal-enrichment of the wind in the Wolf-Rayet stage
which is generally assumed to remain only in the hot SWB for a long time affects
the observationally discernible abundance of the \HII gas.
By these models \cite{kroe06b} could prove for the first time that the
metal release by Wolf-Rayet stars can be mixed within short timescales from the
hot SWB into the warm ionized gas and should become observationally accessible.
As the extreme case for the 85 \Msun star we found a 22\% enhancement of Carbon,
but neglible amouts for N and O.

Since the occurrence of a WR phase is strongly metal dependent, the
enrichment with C should also depend on the average metallicity. This would
mean that any radial gradient of C abundance of \HII regions in galactic disks
is steeper than that of O. And indeed, \cite{este05} (2005) found
d$[\log$(C/O)]/dr = $- 0.058 \pm 0.018 \mbox{ dex kpc}^{-1}$
for the Galactic disk.

\noindent
7) As expected from the distribution of \HII gas the radially projected
\Ha brightness shows a decrease to the center and a slight brightening to the
limb but not as strong as expected according to the increase of the line-of-sight
with impact parameter (\cite{frey03}). This effect depends on the bubble
age and evolves from central brightning to a moderate central trough.
It also demonstrates both:
the neglection of heat conduction and the homogeneous initial density
do not allow a sufficient brightning of heat conductive interfaces so that,
secondly, only the photo-evaporated backflow can contribute to the \Ha
luminosity in present models.
In reality, condensations which become embedded into the hot SWB are
exposed to heat conduction.

\noindent
8) The sweep-up of the slow red supergiant wind by the fast Wolf-Rayet
wind produces remarkable morphological structures and emission signatures which
agree well with observed X-ray luminosity and temperature as well as with the
limb brightening of the radially projected X-ray intensity profile
(for details see \cite{frey06}).

\noindent
9) Connected to 1) the higher compression of the pushed \HII region leads to
a stronger recombination and, by this, a higher energy loss by means of
collisionaly excited line emission. This means that the energy transfer
efficiencies \eps's for both radiative as well as kinetic energies remain
much lower than analytically derived (more than one order of magnitude)
and amount to only a few per mil (\cite{hens07}).
There is almost no dependence on the stellar mass what contrasts expectations
because the energy impact by Lyman continuum photons
and by wind luminosity increase with stellar mass.
Vice versa, since the gas compression is stronger by a more energetic wind
also the energy loss by radiation is more efficient.

\section{Caveats and Required Developments for the Art of Modelling in the
21$^{st}$ Century}

\begin{figure}
\begin{center}
 \includegraphics[width=5.0in]{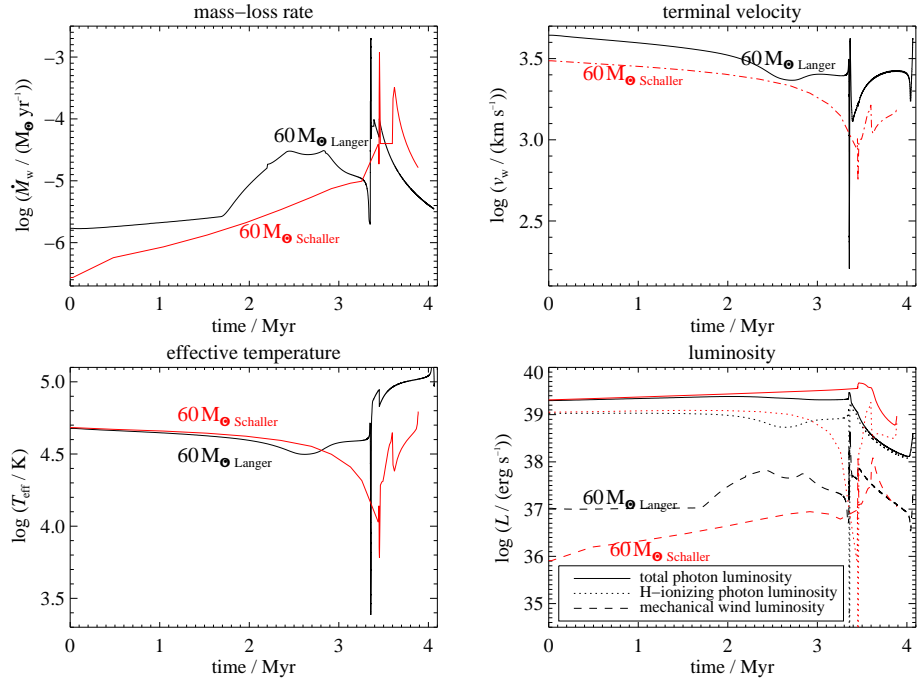}
  \caption{Comparison of the 60 \Msun star parameters from evolutionary models
by Langer \cite{garc96a} and Schaller et al. (1992) (see line notation). 
Notice that in the lower-left figure total luminosity by Schaller et al. 
is represented by the upper curve. (from Kroeger 2006)}
\end{center}
\end{figure}

Nonetheless, words of caution and unfortunately of discouragement have to be
expressed here with respect to various aspects:

At first, the stellar evolutionary models are not yet unique but depend on the authors. In order to get a quantifiable comparison
of the models by Garc\'{\i}a-Segura et al. (1996a, 1996b)
with our 2D radiation-hydrodynamical
simulations (Freyer et al. 2003, 2006) for the 35 and 60 \Msun studies we used
the same stellar parameters. Since no stellar
parameters were available from the same group for the 15 and 85 \Msun models
we had to make use of the Geneva models (\cite{scha92}).
A comparison of the age-dependent  parameters between the Langer
and the Geneva 60 \Msun models (fig. 3) has revealed
enormous differences in the energetics by almost one order of magnitude as
well as that the Wolf-Rayet and Luminous Blue Variable stages occur in
contrary sequence, respectively (\cite{kroe06a}).

Secondly, as presented by George Meynet (this conference) the role of stellar
rotation was until recently totally underestimated, mainly, because of a
lack of model insights. In principle, the radiation-driven stellar wind is
smaller at the equator of a rotating star with respect to the poles
due to van Zeippel's theorem, while in contrast the centrifugal-driven
amount should increase. In addition, the radial and azimuthal redistribution
of fresh fuel and already burned material in the stellar interior is
uncertain as long as at least 2D stellar evolutionary models are lacking.

Hirschi et al. (2004, 2005) showed that taking stellar rotation
into account in the Geneva stellar evolution code increases the yields
for heavy elements by a factor of $1.5 - 2.5$ for stars between $15
\mbox{ and } 30 \mbox{ M}_{\odot}$. For the more massive stars rotation
raises the yields of $^4\mbox{He}$ and other H-burning products like
$^{14}\mbox{N}$, but the yields of He-burning products like
$^{12}\mbox{C}$ are smaller. Additionally, for stars with $M \gtrsim
60 \mbox{ M}_{\odot}$, the evolution differs from that of non-rotating
stars by the following manner: Rotating stars enter the WR regime already
in the course of their main-sequence. Nevertheless, the WR phase has to be
treated in the context of fully self-consistent evolutionary models with
meridional circulations.

Already in non-rotating models for metallicities less than solar two
effects reduce the heavy element release by WR stars: First, the lower the
metallicity the more massive a star has to be to evolve through the WR
stages. Therefore, the number of WR stars decreases with decreasing
metallicity. Schaller et al. (1992) found that with metallicity $Z=0.001$
the minimal initial
H-ZAMS mass for a WR star is $> 80 \mbox{ M}_{\odot}$. At $Z=0.02$ the
minimal initial mass is $> 25 \mbox{ M}_{\odot}$. Second, the lower
the metallicity the shorter are the WR lifetimes, and not all WR
stages are reached. At solar metallicity WR stars enter all three WR
stages (WNL, WNE, WC), whereas at $Z=0.001$ only the WNL phase is
reached (\cite{scha92}). The WR lifetime of an $85 \mbox{
  M}_{\odot}$ star, e.g., is $t_{WR}= 0.204 \times 10^5 \mbox{ yr}$ at
$Z=0.001$ and $t_{WR}= 4.008 \times 10^5 \mbox{ yr}$ at $Z=0.02$
(\cite{scha92}).

At third, the quiescent SF as a self-regulated process is a widely
accepted concept. The stellar feedback can adapt both signs, positive as a
triggering mechanism in a self-propagating manner like in superbubble
shells vs. negative as self-regulation.
Primarily the correlation between the surface density of disk galaxies'
\HI gas and the vertically integrated SF rate derived from the \Ha flux
(\cite{kenn98}) serves as the best proof of a SF self-regulation.
Such self-regulation is a plausible process due to the energy release mainly
by massive stars as demonstrated by \cite{koep95}, whereas its level of the
SF rate is determined by the deposit of the released energy to the ISM.
While the windless i.e. purely photo-ionizing models and analytical
results e.g. by \cite{lask67} reach about one percent for the energy transfer
efficiency \eps\ (see e.g. \cite{frey03}), the radiation-dirven + wind-blown
models by us fall short by more than one order of magnitude even to below
0.1 percent.

As a further aspect, the two-dimensionality of the numerical treatment
must be overcome to 3D in order to allow for a proper description of
turbulent eddies and small-scale inhomogeneities. In addition this requires
also a change in the spatial resolution
of the numerical code: the nested-grid strategy has to be changed to a
flexible mesh adaptivity (adaptive mesh refinement: AMR).
This approach was already developed by Rijkhorst et al. (2006) for the
publicly available and widely used AMR code FLASH.

Last but not least, massive stars are born in OB associations with separations
such low that their wind-driven \HII regions should overlap. For observations
these colliding SWBs should produce a higher X-ray luminosity than expected from
individual massive SWBs. Recent X-ray observations of hot gas even in the Orion
complex (\cite{gued07}) support this expectation. The modelling of such
star-forming regions is highly complex, necessarily 3D, and requires inherently
AMR. Only those explorations will enable us to answer the problem of timescales
and energetics necessary for the observed gas evacuation of star clusters
(\cite{baum08}).

As one recognizes we are still in the natal phase of understanding and modelling
structure and evolution of massive stars and their influence on the ambient ISM.
Much before the end of this 21$^{st}$ century revolutionary observational results
and numerical models will enlighten our present-day ignorance.

\begin{acknowledgments}
The author is gratefully acknowledging contributions and discussions by
Tim Freyer, Joachim K\"oppen, Danica Kroeger, Simone Recci,
Wolfgang Vieser, and Harald W. Yorke.
Part of the work was funded by the DFG under grants HE~1487/17.
G.H. cordially thanks the organizers for the invitation to
this conference. The participation was made possible by grants from the
conference and from the University of Vienna.
\end{acknowledgments}

\begin{discussion}

\discuss{Langer}{To what extent do you need many different stellar
evolution input models for your bubble models; it looked as if the
energy transfer efficiencies in the 4 very different cases you tried
were very similar.}

\discuss{Hensler}{That is correct. From our models the energy
transfer efficiency seems not to depend on the stellar mass although
the power of radiative and wind releases do. This can be understood,
because the stronger stellar wind at larger masses lead to stronger
compression of surrounding gas, and, by this, to stronger cooling by
collisional excitated emission. Different evolutionary models, however, 
affect other issues. We explored e.g. that differences in the kinetic 
energies of the stellar wind (see fig. 3) change the dynamical structure
of the 60 \Msun model in the sense that the fingure-like structures 
(see Freyer et al. 2003) are less pronounced for Schaller et al. (1992)
parameters. In addition, differences in stellar evolutionary models 
also change the LBV-WR sequence and the element release and they have 
stronger effects on the chemical evolution and the self-enrichment 
of HII regions.}

\end{discussion}

\end{document}